\documentclass{article}        
\usepackage{PRIMEarxiv}        

\usepackage[utf8]{inputenc} 
\usepackage[T1]{fontenc}    
\usepackage{hyperref}       
\usepackage{url}            
\usepackage{booktabs}       
\usepackage{amsfonts}       
\usepackage{nicefrac}       
\usepackage{microtype}      
\usepackage{lipsum}
\usepackage{graphicx}
\graphicspath{{media/}}     
\usepackage{orcidlink}
\usepackage{caption}
\usepackage{subcaption}
\usepackage{cite}
\usepackage{array}
\usepackage{multirow}
\usepackage{etoolbox}
\usepackage[linesnumbered,ruled,vlined]{algorithm2e}
\usepackage{amsmath,amssymb,amsfonts}%
\usepackage{amsthm}%
\usepackage{mathrsfs}%
\usepackage[title]{appendix}%
\usepackage{xcolor}%
\usepackage{textcomp}%
\usepackage{manyfoot}%
\usepackage{algorithmicx}%
\usepackage{listings}%
\usepackage{float} 

\title{Leadership Detection via Time-Lagged Correlation-Based Network Inference\thanks{Preprint – This version is not peer-reviewed.}}

\author{
  \orcidlink{0000-0003-2535-3024}Thayanne França da Silva, \orcidlink{0000-0002-5932-9818}Jos\'e Everardo Bessa Maia \\
  Universidade Estadual do Cear\'a - UECE, 60714-903, Fortaleza-CE, Brasil \\
  \texttt{thayanne.silva@aluno.uece.br, jose.maia@uece.br} \\
}

\begin{document}
\maketitle

\begin{abstract}
Understanding leadership dynamics in collective behavior is a key challenge in animal ecology, swarm robotics, and intelligent transportation. Traditional information-theoretic approaches, including Transfer Entropy (TE) and Time-Lagged Mutual Information (TLMI), have been widely used to infer leader-follower relationships but face critical limitations in noisy or short-duration datasets due to their reliance on robust probability estimations. This study proposes a method based on dynamic network inference using time-lagged correlations across multiple kinematic variables: velocity, acceleration, and direction. Our approach constructs directed influence graphs over time, enabling the identification of leadership patterns without the need for large volumes of data or parameter-sensitive discretization. We validate our method through two multi-agent simulations in NetLogo: a modified Vicsek model with informed leaders and a predator-prey model featuring coordinated and independent wolf groups. Experimental results demonstrate that the network-based method outperforms TE and TLMI in scenarios with limited spatiotemporal observations, ranking true leaders at the top of influence metrics more consistently than TE and TLMI.

\end{abstract}


\section{Introduction}

Collective movement is a phenomenon observed both in nature and in artificial systems. Emergent collective behavior arises from local interactions among individuals and can be seen in systems ranging from fish schools, ant colonies, and bird flocks to vehicle traffic and robotic swarms. One key aspect of these systems is the presence of leaders, entities capable of influencing others’ behavior through spatial proximity, signaling, or informed decision-making \cite{fu2024informed}.

Identifying leader-follower patterns helps to understand how groups are organized, improves the coordination of swarm systems, and supports the development of models that explain how decisions are made collectively. However, one of the main challenges lies in distinguishing genuine influence from mere spatial and temporal coincidences, particularly in scenarios where data are noisy, sparse, or derived from indirect observations \cite{basak2020information,pilkiewicz2024mutual}.

Among the most widely used techniques in this domain are the information theoretical approaches: Transfer Entropy (TE) and Time-Lagged Mutual Information (TLMI). TE quantifies the flow of information from one entity to another and has been successfully applied to the movement of animals, people, and vehicles \cite{africa2024traffic,basak2020information}. However, it is sensitive to noise and requires large amounts of data. The TLMI measures the statistical dependence between two-time series at different time lags, helping to reveal delayed interactions. Despite its usefulness, TLMI has several limitations: it requires discretization of continuous variables, it is highly dependent on the choice of lag values, its symmetric nature complicates the inference of directional influence, it is sensitive to small sample sizes, and it may fail to detect temporal dependencies in short or noisy trajectories \cite{pilkiewicz2024mutual,daftari2024entropy}.

This work proposes an alternative approach based on dynamic network inference. This method builds directed graphs that evolve to model influence relationships. Unlike traditional metrics, our approach focuses on temporal sequences of local interactions without requiring probability distribution estimation. We apply it to two NetLogo-based simulations: a modified Vicsek model and a leader-follower agent system with partial independence, comparing results against TE and TLMI.

The main contributions of this work are:
\begin{itemize}
    \item We propose a method for leader-follower detection based on time-lagged correlations across multiple kinematic variables, avoiding the need for probabilistic estimations or large datasets.
    \item We conduct a comparative evaluation against two classical information-theoretic methods, Transfer Entropy and Time-Lagged Mutual Information, demonstrating the superior performance of our method in scenarios with limited spatiotemporal data.
\end{itemize}

The remainder of the paper is structured as follows: Section 2 reviews literature methods for leader-follower detection. Section 3 presents the proposed network inference method. Section 4 describes the simulations and the experimental setup. Section 5 presents the results and comparative analysis. Finally, Section 6 discusses the results, and Section 7 concludes the study with future work.

\section{Review Literature}


Transfer entropy is a widely used metric to analyze potential cause-effect relationships in complex systems \cite{schreiber2000,lizier2008}. It measures how much the past of a process $Y$ helps to predict the future of another process $X$, given that we already know the history of $X$. For discrete processes $X_t$ and $Y_t$, this measure is defined by the following equation:

\begin{equation} TE_{Y \to X} = \sum_{x_t, x_{t-1}, y_{t-1}} p(x_t, x_{t-1}, y_{t-1}) \log \left( \frac{p(x_t \mid x_{t-1}, y_{t-1})}{p(x_t \mid x_{t-1})} \right) \end{equation}

This formulation quantifies the amount of information that $Y_{t-1}$ adds about $X_t$, beyond what is already provided by $X_{t-1}$. It is particularly useful for identifying directional influence and has been applied across domains such as biological systems \cite{daftari2024entropy} and traffic flow dynamics \cite{africa2024traffic}.

However, transfer entropy has limitations. It requires a substantial volume of data and is highly sensitive to noise and parameter selection, such as the spatial cutoff distance used to define potential interactions \cite{basak2020information}. Studies using modified Vicsek models have demonstrated the impact of short trajectories and noise on the stability of transfer entropy calculations \cite{basak2020information}.

An alternative is time-lagged mutual information (TLMI), which evaluates dependencies between the past of one process and the present of another while being less sensitive to data scarcity. TLMI is defined as:

\begin{equation} I_\tau(X; Y) = \sum_{x_t, y_{t-\tau}} p(x_t, y_{t-\tau}) \log \left( \frac{p(x_t, y_{t-\tau})}{p(x_t) , p(y_{t-\tau})} \right) \end{equation}

The asymmetry characteristic of leader-follower interactions can be detected when $I_\tau(X; Y) > I_\tau(Y; X)$, implying that the previous state of $Y$ provides more insight into the future state of $X$ than the reverse. Recent work has shown that TLMI can reveal a local peak in $\tau$ representing the follower’s reaction delay \cite{pilkiewicz2024mutual,daftari2024entropy}.

Despite its popularity, TLMI presents some limitations that may compromise its effectiveness in complex scenarios, such as relies on accurate estimation of joint and marginal probability distributions, not provide directional information, dependence on the selection of an appropriate lag, and the computational cost of TLMI increases rapidly with dimensionality and time-series length, especially when using kernel-based or binning estimation methods \cite{kraskov2004,faes2011,frenzel2007}. 

In addition to these information-theoretic metrics, network-based inference techniques have emerged as powerful tools for uncovering leadership dynamics. For example, \cite{corzo2023sharks} inferred hierarchical social structures among sharks using acoustic tracking data and directed acyclic graphs (DAGs). Similarly, \cite{lopez2023megafauna} applied time-lag analyses to derive leader-follower networks in manta rays based on acoustic presence data.

Other works have leveraged graph-based models and temporal constraints. For instance, \cite{li2012cows} analyzed GPS data from cattle herds to detect local influence patterns via graph sequences, highlighting how velocity correlations and neighborhood constraints can capture transient leadership. Meanwhile, in biological modeling, the Cucker–Smale model with informed leaders has been used to simulate the role of hierarchical leadership in group decision-making and cohesion \cite{fu2024informed}.

In transportation systems, \cite{africa2024traffic} demonstrated the utility of transfer entropy and Granger causality to detect leader-follower behavior among vehicles. Their findings show that temporal lags in lateral movements and acceleration reflect short-lived yet meaningful leadership patterns in mixed-traffic environments.

Collectively, these approaches demonstrate that while information-theoretic metrics offer valuable insights, there remains a methodological gap for interpretable and adaptable techniques. Network-based methods, especially those incorporating spatiotemporal constraints and dynamic topologies, represent a direction for studying leadership in real-world and synthetic systems.

\section{Proposed Method}

This paper proposes a directed network inference-based method to detect leader-follower relationships among mobile agents using trajectory data. The idea of this approach is that a potential leader's past movement can influence another agent's future movement, allowing us to infer directional and temporal dependencies between pairs of agents.

For each pair of agents $(i, j)$, we extract their time series of kinematic variables over a sliding window of size $W$, using a fixed time lag $\tau$. The variables considered include the components of the velocity vector $(v_x, v_y)$, scalar speed $v$, acceleration $a$, and movement direction $\theta$. To evaluate the influence of agent $j$ on agent $i$, we compare the time-lagged series of agent $j$, denoted $s_j(t)$, with the corresponding current series of agent $i$, $s_i(t+\tau)$.

The degree of influence is quantified using the Pearson correlation coefficient between each pair of time-lagged series:

\begin{equation}
\rho_{ij}^{(k)} = \text{corr}(s_j^{(k)}(t), s_i^{(k)}(t+\tau))
\end{equation}

where $k \in \{\text{vx}, \text{vy}, \text{vel}, \text{acc}, \text{dir} \}$. If $\rho_{ij}^{(k)} > \theta$, with $\theta$ being a predefined threshold (e.g., $\theta = 0.85$), the variable is considered indicative of influence from agent $j$ to agent $i$ in that window. The sum of all significant correlations across variables is then used to compute a cumulative influence weight:

\begin{equation}
w_{ij} = \sum_{k \in \text{vars}} \rho_{ij}^{(k)} \cdot \mathbb{1}[\rho_{ij}^{(k)} > \theta]
\end{equation}

If $w_{ij} > 0$, the influence is recorded in the adjacency matrix $A \in \mathbb{R}^{n \times n}$, where the entry $A[i, j] = w_{ij}$ represents a directed edge from $i$ to $j$. This convention is adopted so that summing each row of the global matrix reflects how frequently an agent appeared as a leader across windows.

This process is repeated over all sliding time windows, generating a sequence of dynamic interaction matrices $A_t$. These matrices are then summed to build a cumulative network of inferred interactions:

\begin{equation}
A_{\text{total}} = \sum_{t=0}^{T} A_t
\end{equation}

From this cumulative matrix, we compute centrality scores to estimate each agent's role in the interaction network:
\begin{itemize}
    \item \textbf{Out-degree centrality}: $C_{\text{out}}(i) = \sum_j A_{\text{total}}[i, j]$, representing how often agent $i$ acted as a leader.
    \item \textbf{In-degree centrality}: $C_{\text{in}}(i) = \sum_j A_{\text{total}}[j, i]$, representing how often agent $i$ acted as a follower.
\end{itemize}

The final network structure enables direct visualization of the group's collective behavior, where nodes represent agents and edges represent inferred leader-follower links. The thickness of edges is proportional to the frequency of influence over time. The inference process is formalized in Algorithm~\ref{alg:lf_inference}.

\begin{algorithm}[htbp]
\caption{Inference of Leader-Follower Relationships via Time-Lagged Correlation}
\label{alg:lf_inference}
\KwIn{%
    Trajectory data $D = \{(x_t^i, y_t^i, v_t^i, a_t^i, \theta_t^i)\}$ for each agent $i$ at time $t$\\
    Parameters: Window size $W$, time lag $\tau$, correlation threshold $\theta$
}
\KwOut{%
    Accumulated leadership matrix $A_{\text{total}}$
}
Initialize $A_{\text{total}} \gets 0 \in \mathbb{R}^{n \times n}$ \tcp*{$n =$ number of agents}

\ForEach{time window $[t, t+W+\tau)$}{
    Extract windowed data $D_{\text{window}}$\;
    Initialize matrix $A \gets 0 \in \mathbb{R}^{n \times n}$\;

    \ForEach{pair of agents $(i, j)$, with $i \ne j$}{
        Initialize $w_{ij} \gets 0$\;
        \ForEach{motion variable $k \in \{vx, vy, v, a, \theta\}$}{
            Extract $s_j^{(k)}[t : t+W]$ and $s_i^{(k)}[t+\tau : t+W+\tau]$\;

            \If{lengths match and standard deviations $> 0$}{
                Compute $\rho_{ij}^{(k)} \gets \text{corr}(s_j^{(k)}, s_i^{(k)})$\;
                \If{$\rho_{ij}^{(k)} > \theta$}{
                    $w_{ij} \gets w_{ij} + \rho_{ij}^{(k)}$\;
                }
            }
        }
        \If{$w_{ij} > 0$}{
            Set $A[i, j] \gets w_{ij}$ \tcp*{$j$ influenced $i$}
        }
    }
    $A_{\text{total}} \gets A_{\text{total}} + A$
}
\Return{$A_{\text{total}}$}
\end{algorithm}

\section{Experiment}
\subsection{Simulation}

\subsubsection{Simulation of the Modified Vicsek Model in NetLogo}

A simulation environment was implemented in NetLogo \cite{wilensky1999netlogo} to test the proposed method using an adapted version of the Vicsek model \cite{vicsek1995novel}. The Vicsek model has been extensively applied in studies on collective movement, where particles adjust their direction according to neighboring agents, giving rise to emergent dynamics such as fluctuations and phase transitions influenced by noise and density.

In our implementation, the model incorporates variable speeds and differentiated roles (leaders and followers), a standard extension in recent studies on informed leadership and heterogeneous agent influence \cite{fu2024informed}.

The agent population consists of two breeds: leaders ($5$ agents) and followers ($100$ agents). Specifically, agents with IDs $0$ through $4$ are designated as leaders, while agents with IDs $5$ through $104$ act as followers. At initialization, each agent is randomly positioned within the environment and assigned an initial heading and speed. Leaders are red and move with slightly higher initial speeds (between $0.6$ and $1.0$ units per tick), while followers, colored blue, start with speeds ranging from $0.3$ to $0.6$.

Agents interact based on a circular interaction radius of $3$ units. At each time step, agents update their heading based on the average heading of their local neighbors within this radius. A leader heading weight parameter (set to $2$) amplifies the influence of nearby leaders when followers compute their alignment. A small amount of uniform random noise is added to the heading update to simulate natural variability, as initially proposed in the Vicsek model \cite{vicsek1995novel}.

Follower agents dynamically adjust their speed to align with the average speed of nearby leaders using a speed adjustment rate of $0.2$. This reflects followers' tendency to align directionally and adapt their movement pace based on leaders' influence.

We simulate scenarios with limits in NetLogo since scenarios without limits are unrealistic because the agents teleport in the opposite direction, which does not occur in the real scenario. Thus, when an agent reaches the boundary of the environment, it performs a rotational maneuver between $150^\circ$ and $210^\circ$ and then moves in the new direction for $10$ ticks. During this period, the agent is marked as being in a "drift" state, during which it does not update its direction or velocity based on its neighbors, thus avoiding artificial clustering at the boundaries. Additionally, we set the NetLogo environment as a large scenario for the agents compared to the other simulation used in this paper so that the agents have more space to move around.

The simulation was run for a fixed number of time steps, generating trajectory datasets with varied motion dynamics. This setting provides a controlled yet realistic environment to evaluate the performance of information-theoretic and network-based methods in detecting asymmetric interactions between agents.

\subsubsection{Modified Wolf Sheep Predation Simulation}

To further evaluate our proposed leader-follower detection method, we also developed a modified version of the classical Wolf-Sheep Predation model available in NetLogo \cite{wilensky1999netlogo}. 

Our modified simulation introduces a hybrid predator population comprising wolves that hunt independently and wolves that operate in coordination under the command of a designated alpha leader. This setup enables the generation of trajectory data that includes uncoordinated and coordinated group behavior, which is essential for testing the inference of asymmetric interactions in dynamic environments.

The environment has three breeds: sheep, wolves, and a single alpha wolf. At initialization, $100$ sheep are randomly distributed across the environment and move with a simple random walk behavior. A total of $30$ wolves are also created, divided into two distinct groups:

\begin{itemize} 
\item \textbf{Independent wolves} ($15$ agents, IDs $115$ to $129$): These wolves patrol the environment autonomously. Each has an individual chance of identifying and chasing nearby sheep based on a probabilistic hunting trigger. Once a target is selected, the wolf pursues and consumes the prey independently.
\item \textbf{Pack wolves} ($14$ agents, IDs $101$ to $114$) and \textbf{Alpha wolf} ($1$ agent, ID $100$): These wolves follow a single alpha wolf. The alpha wolf randomly explores the environment and occasionally selects a sheep as the target for coordinated hunting. Once a target is set, all pack wolves synchronize their movement direction and increase speed to converge on the prey. The hunt is only successful if the alpha and at least one pack member are simultaneously within a certain radius of the prey, emulating cooperative hunting behavior observed in natural settings \cite{fu2024informed}.
\end{itemize}

To maintain realistic group dynamics, pack wolves update their speed proportionally to the alpha’s speed and adapt their position to remain behind or around the leader. The alpha wolf is responsible for initiating and leading all collective actions, and its influence is explicitly encoded in the simulation logic via centralized target selection and spatial guidance.

The simulation combines independent and leader-driven behaviors within a single group of agents, creating a suitable environment to evaluate the performance of different methods for detecting leader-follower moviment.

\subsection{Experiment setup}

To evaluate the performance of the three leader-follower detection methods, we designed a set of experiments based on two multi-agent simulations implemented in NetLogo: a modified version of the Vicsek model and the Wolf-Sheep Predation model. The table below summarizes the configuration parameters for each method and dataset:

\begin{table}[H]
\centering
\caption{Experimental configuration for each method and simulation.}
\scriptsize
\label{tab:experimental-setup}
\begin{tabular}{|l|l|l|l|}
\hline
\textbf{Aspect} & \textbf{Transfer Entropy} & \textbf{Mutual Information} & \textbf{Network Inference} \\ \hline
\textbf{Simulation 1} & \multicolumn{3}{c|}{Wolf-Sheep Predation (15 independent, 14 follower and 1 alpha)} \\ \hline
\textbf{Simulation 2} & \multicolumn{3}{c|}{Vicsek Model (5 leaders, 100 followers)} \\ \hline
\textbf{Features used} & \multicolumn{3}{c|}{Velocity, Acceleration, Direction} \\ \hline
\textbf{Simulation time of simulation 1} & \multicolumn{3}{c|}{500 steps} \\ \hline
\textbf{Simulation time of simulation 2} & \multicolumn{3}{c|}{1000 steps} \\ \hline
\textbf{Window size of simulation 1} & \multicolumn{3}{c|}{50 steps (10\% of simulation time)} \\ \hline
\textbf{Window size of simulation 2} & \multicolumn{3}{c|}{100 steps (10\% of simulation time)} \\ \hline
\textbf{Lag range} & Embedding = 1 & [-5, +5] & Fixed lag = 1 \\ \hline
\textbf{Threshold} & $\Delta$TE $>$ 0.2 & MI $>$ 0.3 & Pearson $r > 0.85$ \\ \hline
\textbf{Discretization (if any)} & No & 5 bins (equal-width) & No \\ \hline
\end{tabular}
\end{table}

The window sizes of $50$ (for the Wolf-Sheep simulation) and $100$ (for the Vicsek model) were chosen following the recommendation by \cite{Trumler}, who suggest configuring the maximum time window for collective motion detection as $10\%$ of the total simulation duration. This ensures that each analysis window captures a sufficiently long sequence of interactions to reveal temporal dependencies while preserving the dynamics' resolution. 

The lag ranges $[-5, +5]$ used in the mutual information method capture possible delayed responses between agents. In the case of Transfer Entropy, we used the standard embedding of $1$ and $k=4$ neighbors, consistent with prior studies. A Pearson correlation threshold of $0.85$ was selected for the network inference approach to ensure that only strong and reliable directional relationships were considered. Discretization into five bins was applied using the TLMI method to reduce noise while preserving feature variability. All approaches used the same features, velocity, acceleration, and direction, to maintain consistency across methods.

\section{Results}

\subsection{Modified Wolf Sheep Predation}

To evaluate the performance of our network inference method in detecting leader-follower relationships, we conducted experiments using a modified version of the Wolf Sheep Predation model, in which an alpha leader (agent $100$) is followed by $14$ agents (IDs $101$–$114$), while the remaining $15$ agents (IDs $115$–$129$) move independently in the hunt. Each simulation was run for $500$-time steps in $10$ independent runs, and the ranking of the true leader (agent $100$) was recorded according to three different metrics: network inference, TE, and TLMI (Table ~\ref{tab:ranking_lider}).

\begin{table}[ht]
\centering
\scriptsize
\caption{Ranking of the top 5 agents (agents 0–4) across 10 distinct simulations of the Modified Wolf Sheep Predation simulation for each leadership detection metric.}
\label{tab:ranking_lider}
\begin{tabular}{c|c|c|c}
\hline
\textbf{Simulation} & \textbf{Our approach} & \textbf{TE} & \textbf{TLMI} \\
\hline
1  & 1°  & 26° & 13° \\
2  & 1°  & 4°  & 16° \\
3  & 1°  & 16° & 12° \\
4  & 1°  & 13° & 2°  \\
5  & 1°  & 11° & 16° \\
6  & 1°  & 10° & 3°  \\
7  & 1°  & 22° & 14° \\
8  & 1°  & 21° & 3°  \\
9  & 1°  & 12° & 16° \\
10 & 1°  & 15° & 4°  \\
\hline
\textbf{Top 1}         & $10$x $\rightarrow 100\%$  & $0$x $\rightarrow 0\%$            & $1$x $\rightarrow 10\%$           \\
\textbf{Top 3}         & $10$x $\rightarrow 100\%$ & $0$x $\rightarrow 0\%$            & $3$x   $\rightarrow 30\%$        \\
\textbf{Top 5}         & $10$x $\rightarrow 100\%$ & $1$x $\rightarrow 10\%$            & $4$x   $\rightarrow 40\%$        \\
\hline
\end{tabular}
\end{table}

The results demonstrate a advantage of the network inference approach in consistently identifying the alpha leader. In all $10$ simulations, the proposed method ranked agent $100$ in the top position, demonstrating its robustness in capturing the asymmetric and structured influence exerted by the leader on its followers. This performance demonstrates the method's sensitivity to structured influence patterns and ability to robustly distinguish asymmetric interactions characteristic of leadership dynamics, even in short trajectory windows and under partial coordination. In contrast, the transfer entropy method failed to identify the leader in the top $5$ in most simulations, ranking agent $100$ only once in the top 5 and never in the top $3$. TLMI performed slightly better than TE, identifying the leader in the Top 3 in $3$ simulations and the Top 5 in $4$, but it failed to achieve consistent detection. 

These results suggest that network inference is more efficient and stable under the constraints of short spatiotemporal series. The simulation length of only 500 time steps may have limited the performance of the transfer entropy and mutual information methods, which typically require larger observation windows to estimate directional or time-dependent dependencies reliably. These approaches are also more noise-sensitive and lack sufficient historical data to capture lagged interactions. These factors can obscure influence patterns in leader-follower dynamics over short periods. 

\begin{figure}[ht]
    \centering

    \begin{subfigure}[b]{.9\textwidth}
        \centering
        \includegraphics[width=\textwidth]{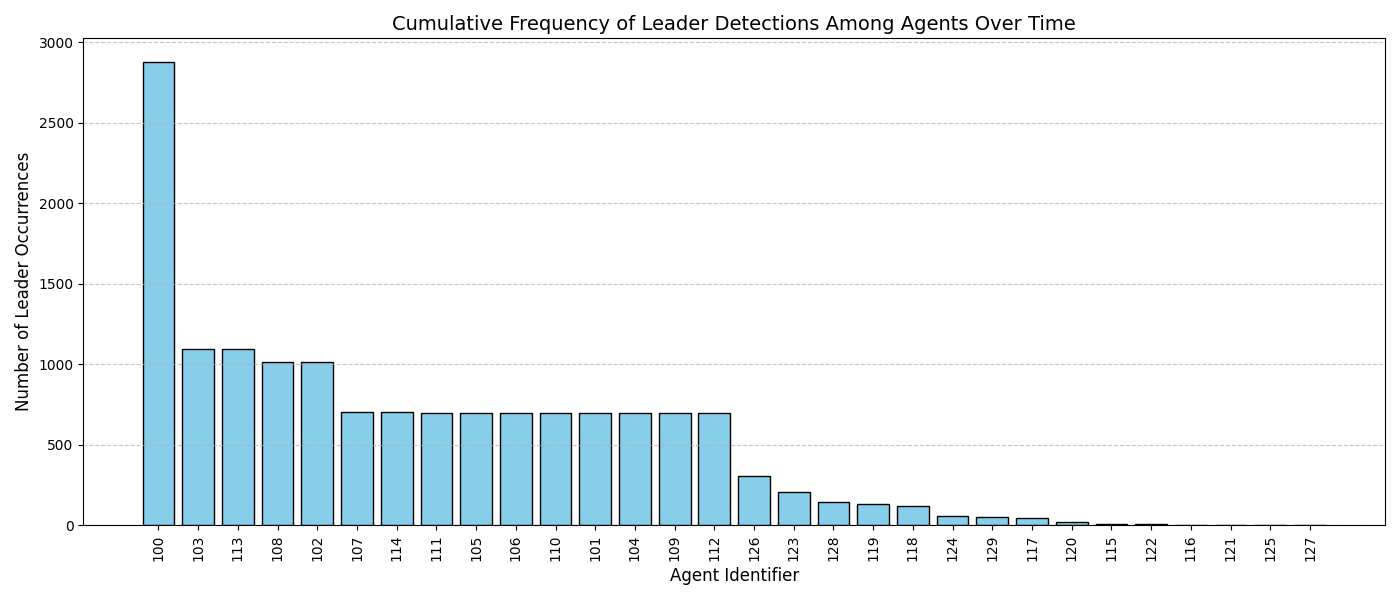}
        \caption{Network Inference (proposed approach)}
        \label{fig:rede-lobos-lider}
    \end{subfigure}

    \vspace{0.3cm}

    \begin{subfigure}[b]{0.9\textwidth}
        \centering
        \includegraphics[width=\textwidth]{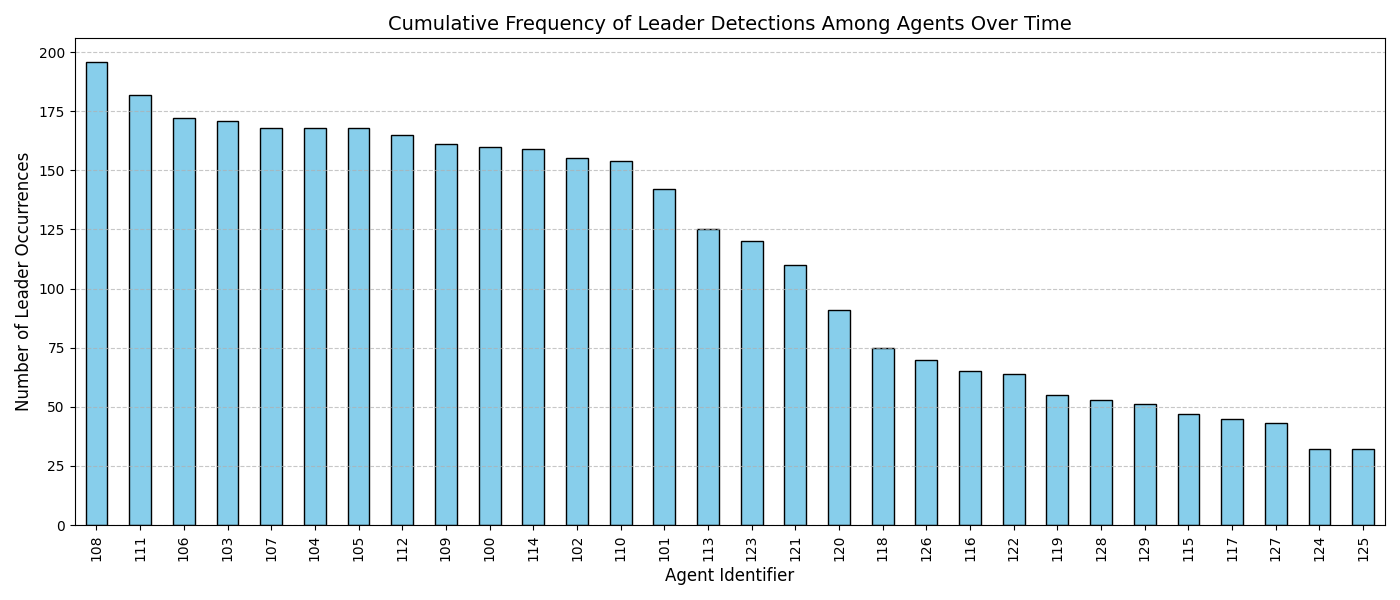}
        \caption{Transfer Entropy}
        \label{fig:entropia-lobo-lider}
    \end{subfigure}

    \vspace{0.3cm}

    \begin{subfigure}[b]{0.9\textwidth}
        \centering
        \includegraphics[width=\textwidth]{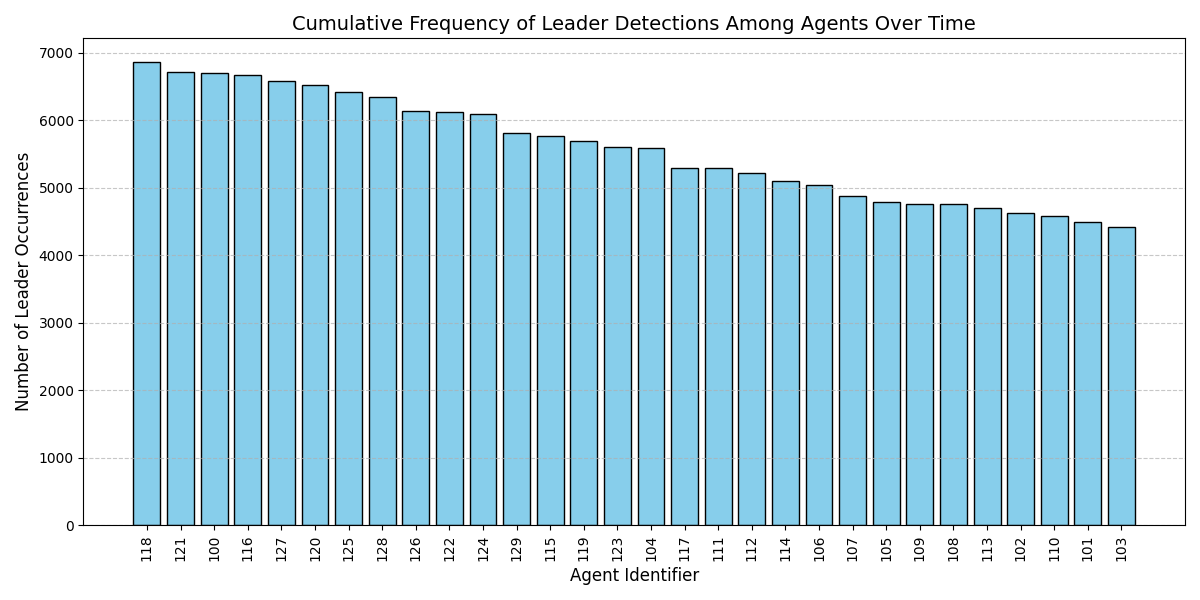}
        \caption{Time-Lagged Mutual Information}
        \label{fig:mi-lobo-lider}
    \end{subfigure}

    \caption{Cumulative frequency of leadership detection for each agent in the Modified Wolf Sheep Predation simulation across the three analyzed approaches from one of the ten conducted simulations: (a) Network inference, (b) Transfer entropy, and (c) Time-lagged mutual information.}
    \label{fig:lobos-lideres}
\end{figure}

To complement the results in the table, Figure~\ref{fig:lobos-lideres} presents the cumulative frequency of leadership detection for each agent in one of the $10$ simulations of the modified Wolf Sheep Predation model in NetLogo, comparing the three approaches. In this simulation, agent $100$ is the alpha wolf, agents with IDs $101$ through $114$ are the follower wolves, and agents with IDs $115$ through $129$ are the independent wolves.

It is observed that the network inference-based approach (Figure~\ref{fig:rede-lobos-lider}) was the only one capable of accurately capturing the hierarchical structure present in the collective dynamics. The leader agent (ID $100$) was consistently the most frequently detected as a leader over time, followed by the subordinate agents (direct followers), and finally, the independent agents, who exhibited a low detection frequency. This ordering highlights the sensitivity of network inference to the organizational structure of the Wolf Sheep Predation simulation, as it not only captured direct leadership but also recognized levels of influence resulting from relationships with followers of independent agents unrelated to the leader.

In contrast, the approaches based on transfer entropy (Figure~\ref{fig:entropia-lobo-lider}) and time-lagged mutual information (Figure~\ref{fig:mi-lobo-lider}) failed to reflect the underlying hierarchical structure. Both produced a relatively homogeneous distribution across agents, with no evident prominence for the true leader. This behavior may be because these techniques require longer time series to converge toward significant dependency patterns. Given that the analyzed data comprise only $500$ simulation steps, these approaches showed limited performance.

These results demonstrate that our proposed approach based on network inference outperformed in scenarios with sparse spatiotemporal data or limited observation time. Furthermore, the ability to distinguish true leaders from their followers in dynamic environments shows its application in problems such as collective behavior monitoring, social influence analysis, and distributed coordination in robotics.

\subsection{Modified Vicsek Model}

Table~\ref{tab:ranking_lider_vicsek} presents the position of the five leading agents (IDs $0$ to $4$) in $10$ independent simulations using the three approaches. All approaches were evaluated under the same modified Vicsek model configuration, with $5$ leaders and $100$ followers, simulated for $1000$ time steps.

\begin{table}[htbp]
\centering
\scriptsize
\caption{Ranking of the top 5 agents (agents 0–4) across 10 distinct simulations of the Modified Vicsek simulation for each leadership detection metric.}
\label{tab:ranking_lider_vicsek}
\begin{tabular}{c|c|c|c}
\hline
\textbf{Simulation} & \textbf{Our approach} & \textbf{TE} & \textbf{TLMI} \\
\hline
1  & 4: 1°, 3: 2°, 1: 3°, 0: 4°, 2: 5°  &  2: 18°, 1: 21°, 0: 23°, 3: 26°, 4: 28° &  0: 21°, 3: 29°, 1: 36°, 4: 39°, 2: 50°\\
2  & 4: 1°, 1: 2°, 0: 3°, 2: 94°, 3: 95°  & 2: 8°, 0: 13°, 4:20°, 3: 22°, 1:24°  &  4: 7°, 0: 43°, 1: 48°, 3: 52°, 2: 60° \\ 
3  & 0: 1°, 1: 2°, 4: 3°, 2: 4°, 3: 96°  & 3: 13°, 2: 15°, 1: 16°, 4: 17°, 0: 19° &  3: 44°, 1: 54°, 0: 60°, 4: 74°, 2: 79°\\
4  & 1: 3°, 2: 9°, 4: 10°, 0: 87°, 3: 92°  & 3: 11°, 4: 26°, 1: 27°, 0: 29°, 2: 32° &  3: 13°, 1: 14°, 2: 35°, 4: 60°, 0: 67° \\
5  & 3: 2°, 2: 5°, 0: 89°, 4:90°, 1: 98°  & 1: 13°, 3: 28°, 2: 39°, 0: 42°, 4: 43° &  2: 3°, 4: 7°, 1: 16°, 0: 38°, 3: 82°\\
6  & 0: 1°, 3: 2°, 1: 3°, 4: 8°, 2: 35°  & 2: 16°, 4: 20°, 0: 21°, 3: 22°, 1: 23° &   3: 4°, 1: 12°, 0: 15°, 2: 85°, 4: 95°\\
7  & 1: 2°, 3: 4°, 2: 10°, 4: 21°, 0: 99°  & 3: 7°, 0: 10°, 4: 15°, 2: 28°, 1: 32° &  3: 28°, 0: 29°, 1: 58°, 2: 73°, 4: 87°\\
8  & 1: 1°, 4: 2°, 0: 4°, 2: 36°, 3: 98°  & 0: 7°, 4: 10°, 3: 12°, 1: 13°, 2: 15° &   2: 12°, 1: 17°, 3: 41°, 0: 46°, 4: 47°\\
9  & 1: 1°, 0: 2°. 3: 4°, 4: 5°, 2: 14°  & 3: 11°, 1: 13°, 2: 14°, 0: 22°, 4: 26° &  4: 4°, 0: 8°, 1: 26°, 2: 48°, 3: 85°\\
10 & 1: 1°, 0: 14°, 2: 18°, 4: 22°, 3: 30°  & 3: 19°, 2: 28°, 1: 30°, 4: 36°, 0: 37° &  3: 13°, 0: 16°, 1: 21°, 4: 37°, 2: 75°\\
\hline
\textbf{Top 5}         & $28$x $\rightarrow 56\%$ & $0$x  $\rightarrow 0\% $ &  $3$x  $\rightarrow 6\% $ \\
\textbf{Top 10)}         & $32$x $\rightarrow 64\%$  & $5$x  $\rightarrow 10\%$ & $6$x  $\rightarrow 12\% $ \\
\hline
\end{tabular}
\end{table}

The network-based method demonstrated superior performance in identifying expected leaders compared to TE and TLMI. In $28$ of the $50$ cases evaluated, the method placed the leaders among the top five positions, resulting in $56\%$ coverage. In $64\%$ of cases ($32$ out of $50$), leaders appeared in the top ten, indicating that the approach remains reliable even when leaders are not ranked at the top. 

In contrast, the TE showed low sensitivity to the underlying leadership structure. None of the positions evaluated appeared in the top five, and only five cases ($10\%$) were ranked in the top ten. The TLMI exhibited an intermediate level of performance, falling between the network inference and TE methods. While TLMI outperformed TE by identifying the leading agents among the top five positions in $6\%$ of the cases and the top ten in $12\%$ of the cases, its performance remained lower than that of the proposed network inference method. These results show that TE and TLMI struggled, likely due to their reliance on accurate estimates of conditional probability distributions, combined with sensitivity to the choice of lag and sample size.

\begin{figure}[htbp]
    \centering

    \begin{subfigure}[b]{0.9\textwidth}
        \centering
        \includegraphics[width=\textwidth]{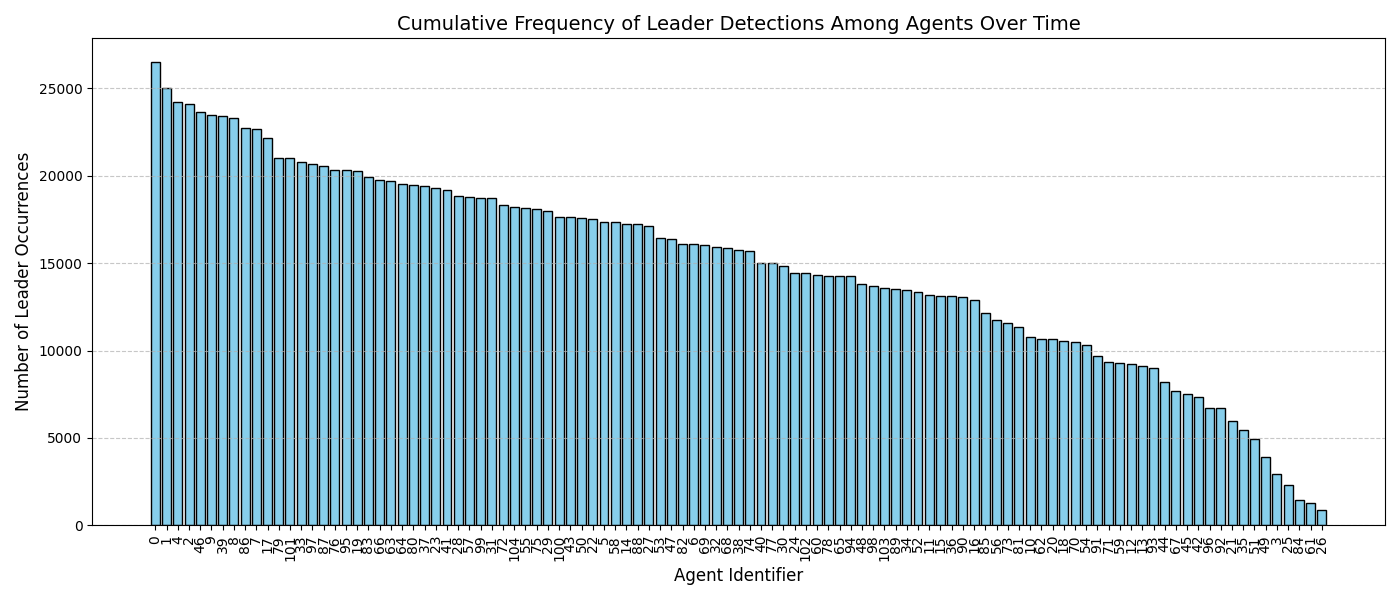}
        \caption{Network Inference (proposed approach)}
        \label{fig:rede-vicsek-lider}
    \end{subfigure}

    \vspace{0.3cm}

    \begin{subfigure}[b]{0.9\textwidth}
        \centering
        \includegraphics[width=\textwidth]{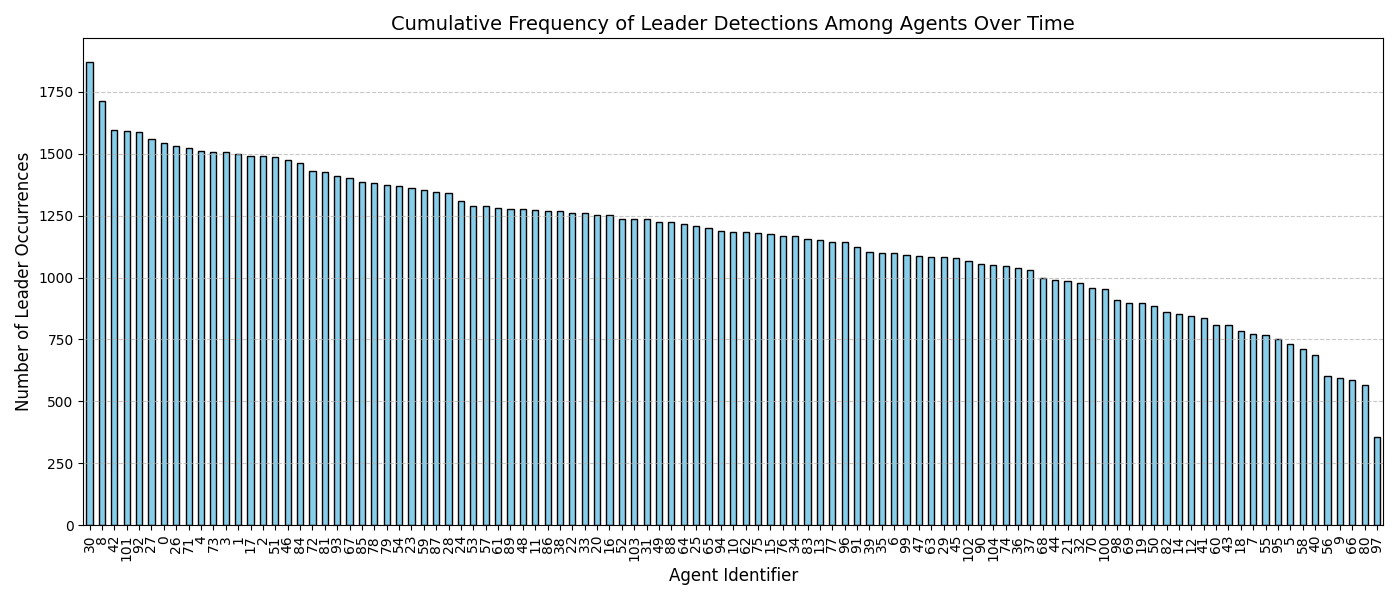}
        \caption{Transfer Entropy}
        \label{fig:entropia-vicsek-lider}
    \end{subfigure}

    \vspace{0.3cm}

    \begin{subfigure}[b]{0.9\textwidth}
        \centering
        \includegraphics[width=\textwidth]{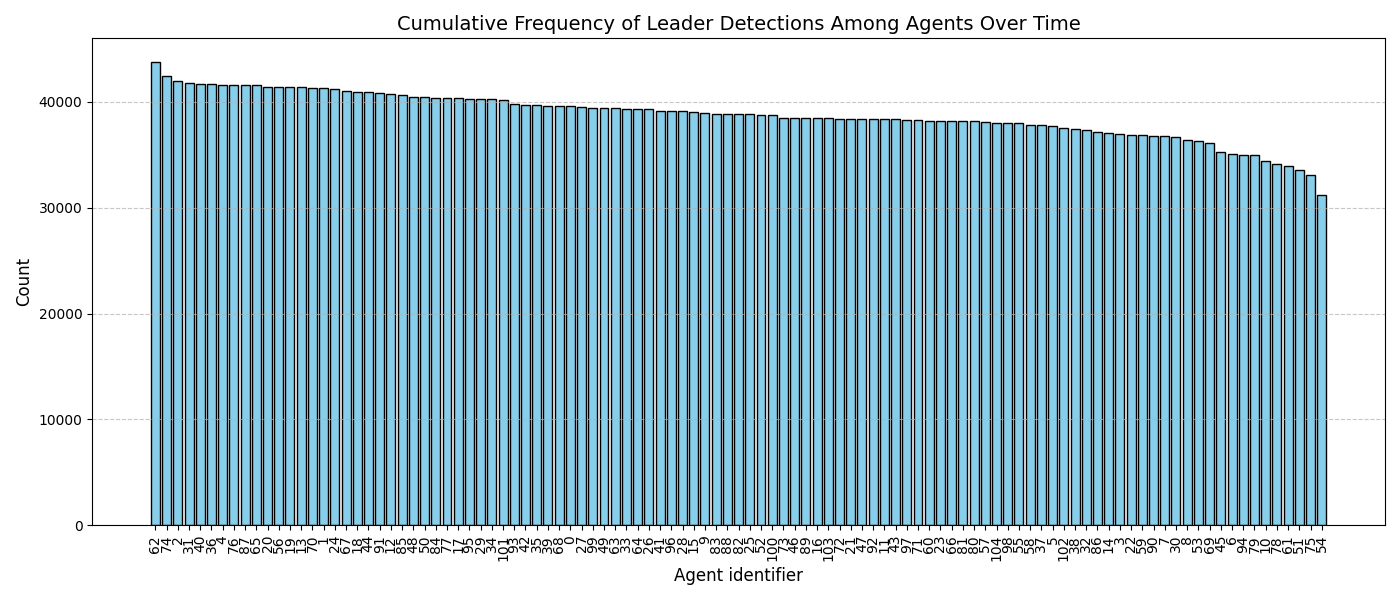}
        \caption{Time-Lagged Mutual Information}
        \label{fig:mi-vicsek-lider}
    \end{subfigure}

    \caption{Cumulative frequency of leadership detection for each agent in the Modified Vicsek simulation across the three analyzed approaches from one of the ten conducted simulations: (a) Network inference, (b) Transfer entropy, and (c) Time-lagged mutual information.} 
    \label{fig:vicsek-lideres}
\end{figure}

These results are also reflected in the histograms of Figures ~\ref{fig:rede-vicsek-lider}, ~\ref{fig:entropia-vicsek-lider}, and ~\ref{fig:mi-vicsek-lider}, which show the cumulative frequency of leadership detection for each agent in one of the $10$ simulations. Figure ~\ref{fig:rede-vicsek-lider}, corresponding to our approach, shows that leader agents (IDs $0$ to $4$) appear among the agents most frequently detected as leaders. In contrast, Figure ~\ref{fig:entropia-vicsek-lider}, corresponding to Transfer Entropy, shows a more dispersed distribution without systematically highlighting the real leading agents. In Figure~\ref{fig:mi-vicsek-lider}, the TLMI distribution is flatter and less polarized. Although some of the actual leader agents appear near the top, many others exhibit similar detection frequencies, indicating difficulty distinguishing leader agents from followers.

Although this simulation lasted longer than the Modified Wolf Sheep Predation scenario, it was still insufficient to enhance the performance of TE and TLMI. In scenarios with limited temporal windows, these methods' ability to capture directional causal relationships may be compromised.

\section{Discussion}

The comparative analysis between the approaches revealed insights into each method's performance, applicability, and limitations for leader-follower detection in trajectory-based simulations. First, the superior performance of the network inference method in both scenarios evaluated (the modified Wolf Sheep Predation and Vicsek models) suggests a advantage in identifying asymmetric influence relationships, especially in environments with structured group dynamics and short spatiotemporal series or limited observation time. 

In the Wolf Sheep Predation scenario, the proposed method consistently classified the alpha leader (agent $100$) as the top leader in all $10$ simulations. In addition to capturing the structure of the environment, where the leader agent appears in the first position, followed by his followers, and then by the independents. In contrast, TE and IMTL showed considerable instability, failing to identify the leader in most simulations.

Similarly, in the Vicsek model with $1000$ time steps, the network inference approach ranked agents $0$–$4$ (true leaders) among the top $5$ in $56\%$ of the cases and the top $10$ in $64\%$. However, TE demonstrated limited effectiveness in this scenario, with only $10\%$ of detections falling into the top 10. TLMI performed slightly better, capturing $6\%$ of the leaders in the top $5$ and $12\%$ of the top $10$, but still significantly underperforming compared to the network-based approach. 

The limited performance of TE and TLMI may be explained by their dependence on the volume of spatiotemporal data. Both techniques require a substantial number of observations to produce reliable results, which becomes a limitation when working with short or fragmented trajectories. In the case of TLMI, the need to define a suitable time lag and to discretize continuous variables can further increase the likelihood of underfitting when data is scarce.

On the other hand, the network inference method uses correlations between motion features, such as speed, acceleration, and heading, measured over time to construct directed dynamic graphs reflecting agent influence. This structure allows leadership patterns to emerge more clearly, even when the data is affected by noise.

However, it is important to note that the superior performance of the network inference method in the Wolf Sheep Predation scenario, characterized by a smaller number of agents and a shorter simulation length of 500-time steps, may be attributed to the simplicity and clarity of the hierarchical structure present in that controlled environment. The method differentiated the alpha leader, direct followers, and independent agents. However, its performance in the modified Vicsek model, which involved a larger number of agents and a longer temporal window, was lower than that of the wolves. This observation suggests our approach is well-suited for scenarios with fewer agents and limited trajectory data. However, its generalizability to more complex and dynamic real-world environments remains to be verified. In such contexts, where trajectories are longer, richer, and more continuous, information-theoretic methods like Transfer Entropy (TE) and Time-Lagged Mutual Information (TLMI) may demonstrate superior performance. Their ability to capture temporal causality and delayed dependencies helps them to detect influence patterns that correlation-based methods might overlook, especially when interactions are indirect or mediated through intermediate agents. It is necessary to evaluate the scalability and limitations of our approach under real-world conditions and to better understand the comparative advantages of each method across diverse scenarios.

\section{Conclusion}

This work presented a new approach to detecting leader-follower relationships using network inference based on time-lagged correlations of kinematic features. Using controlled multi-agent simulations with known leadership structures, the method was evaluated against two classical information-theoretic techniques: Transfer Entropy and Time-Lagged Mutual Information.

The results showed that the network-based method performed better in scenarios with shorter observation periods than traditional techniques, which performed worse in these scenarios. This may be because our approach does not rely on large datasets or complex probability calculations, which makes it a good option for applications with limited data.

In future work, we intend to apply the proposed approach in real scenarios and environments with larger volumes of data to investigate its performance under conditions more favorable to information-theoretic approaches. This comparison will allow a more comprehensive understanding of our technique's limits and potential at different levels of complexity and data availability.

\bibliography{references}

\begin{thebibliography}{16}
\providecommand{\natexlab}[1]{#1}
\providecommand{\url}[1]{\texttt{#1}}
\expandafter\ifx\csname urlstyle\endcsname\relax
  \providecommand{\doi}[1]{doi: #1}\else
  \providecommand{\doi}{doi: \begingroup \urlstyle{rm}\Url}\fi

\bibitem[Africa et~al.(2024)Africa, Quiangco, and Go]{africa2024traffic}
D.~D. Africa, R.~B.~D. Quiangco, and C.~K. Go.
\newblock Lag and duration of leader–follower relationships in mixed traffic using causal inference.
\newblock \emph{Chaos}, 34:\penalty0 013130, 2024.
\newblock \doi{10.1063/5.0166785}.
\newblock URL \url{https://doi.org/10.1063/5.0166785}.

\bibitem[Basak et~al.(2020)Basak, Sattari, Motaleb, Horikawa, and Komatsuzaki]{basak2020information}
U.~S. Basak, S.~Sattari, H.~M. Motaleb, K.~Horikawa, and T.~Komatsuzaki.
\newblock An information-theoretic approach to infer the underlying interaction domain among elements from finite length trajectories in a noisy environment.
\newblock \emph{Physical Review E}, 102\penalty0 (1):\penalty0 012404, 2020.

\bibitem[Daftari et~al.(2024)Daftari, Mayo, Lemasson, Biedenbach, and Pilkiewicz]{daftari2024entropy}
K.~Daftari, M.~L. Mayo, B.~H. Lemasson, J.~M. Biedenbach, and K.~R. Pilkiewicz.
\newblock Probing asymmetric interactions with time-separated mutual information: A case study using golden shiners.
\newblock \emph{Entropy}, 26\penalty0 (775), 2024.
\newblock \doi{10.3390/e26090775}.
\newblock URL \url{https://doi.org/10.3390/e26090775}.

\bibitem[Faes et~al.(2011)Faes, Nollo, and Porta]{faes2011}
L.~Faes, G.~Nollo, and A.~Porta.
\newblock Information-based detection of nonlinear granger causality in multivariate processes via a nonuniform embedding technique.
\newblock \emph{Physical Review E}, 83\penalty0 (5):\penalty0 051112, 2011.
\newblock \doi{10.1103/PhysRevE.83.051112}.
\newblock URL \url{https://doi.org/10.1103/PhysRevE.83.051112}.

\bibitem[Fernández-Gracia et~al.(2024)]{lopez2023megafauna}
J.~Fernández-Gracia et~al.
\newblock Inferring leader-follower behavior from presence data in the marine environment: A case study on reef manta rays.
\newblock \emph{arXiv preprint}, 2024.

\bibitem[Frenzel and Pompe(2007)]{frenzel2007}
S.~Frenzel and B.~Pompe.
\newblock Partial mutual information for coupling analysis of multivariate time series.
\newblock \emph{Physical Review Letters}, 99\penalty0 (20):\penalty0 204101, 2007.
\newblock \doi{10.1103/PhysRevLett.99.204101}.
\newblock URL \url{https://doi.org/10.1103/PhysRevLett.99.204101}.

\bibitem[Fu et~al.(2024)]{fu2024informed}
Y.~Fu et~al.
\newblock Research on group behavior modeling and individual interaction modes with informed leaders.
\newblock \emph{Mathematics}, 12\penalty0 (1160), 2024.
\newblock \doi{10.3390/math12081160}.
\newblock URL \url{https://doi.org/10.3390/math12081160}.

\bibitem[Kraskov et~al.(2004)Kraskov, Stögbauer, and Grassberger]{kraskov2004}
A.~Kraskov, H.~Stögbauer, and P.~Grassberger.
\newblock Estimating mutual information.
\newblock \emph{Physical Review E}, 69\penalty0 (6):\penalty0 066138, 2004.
\newblock \doi{10.1103/PhysRevE.69.066138}.
\newblock URL \url{https://doi.org/10.1103/PhysRevE.69.066138}.

\bibitem[Li and Guibas(2012)]{li2012cows}
Y.~Li and L.~J. Guibas.
\newblock Leader-follower relationships from trajectory data – a case study.
\newblock In \emph{CG:YRF}, 2012.

\bibitem[Lizier et~al.(2008)Lizier, Prokopenko, and Zomaya]{lizier2008}
J.~T. Lizier, M.~Prokopenko, and A.~Y. Zomaya.
\newblock Local information transfer as a spatiotemporal filter for complex systems.
\newblock \emph{Physical Review E}, 77\penalty0 (2):\penalty0 026110, 2008.
\newblock \doi{10.1103/PhysRevE.77.026110}.
\newblock URL \url{https://doi.org/10.1103/PhysRevE.77.026110}.

\bibitem[López(2023)]{corzo2023sharks}
M.~T.~Corzo López.
\newblock Inference of leadership networks of marine megafauna from acoustic data.
\newblock Master's thesis, IFISC – Institute for Cross-Disciplinary Physics and Complex Systems, 2023.

\bibitem[Pilkiewicz et~al.(2024)]{pilkiewicz2024mutual}
K.~R. Pilkiewicz et~al.
\newblock Time-separated mutual information reveals key characteristics of asymmetric leader-follower interactions in golden shiners.
\newblock \emph{bioRxiv}, 2024.
\newblock \doi{10.1101/2024.03.05.583541}.
\newblock URL \url{https://doi.org/10.1101/2024.03.05.583541}.

\bibitem[Schreiber(2000)]{schreiber2000}
T.~Schreiber.
\newblock Measuring information transfer.
\newblock \emph{Physical Review Letters}, 85\penalty0 (2):\penalty0 461--464, 2000.
\newblock \doi{10.1103/PhysRevLett.85.461}.
\newblock URL \url{https://doi.org/10.1103/PhysRevLett.85.461}.

\bibitem[Trumler and Gerdes(2008)]{Trumler}
W.~Trumler and M.~Gerdes.
\newblock Towards an automated detection of self-organizing behavior.
\newblock In \emph{INFORMATIK 2008. Beherrschbare Systeme – dank Informatik. Band 2}, pages 739--746. Gesellschaft für Informatik e. V., Bonn, 2008.
\newblock ISBN 978-3-88579-228-4.

\bibitem[Vicsek et~al.(1995)Vicsek, Czirók, Ben-Jacob, Cohen, and Shochet]{vicsek1995novel}
T.~Vicsek, A.~Czirók, E.~Ben-Jacob, I.~Cohen, and O.~Shochet.
\newblock Novel type of phase transition in a system of self-driven particles.
\newblock \emph{Physical Review Letters}, 75\penalty0 (6):\penalty0 1226--1229, 1995.
\newblock \doi{10.1103/PhysRevLett.75.1226}.
\newblock URL \url{https://doi.org/10.1103/PhysRevLett.75.1226}.

\bibitem[Wilensky(1999)]{wilensky1999netlogo}
U.~Wilensky.
\newblock Netlogo.
\newblock \url{https://ccl.northwestern.edu/netlogo/}, 1999.
\newblock Center for Connected Learning and Computer-Based Modeling, Northwestern University, Evanston, IL.

\end{thebibliography}

\end{document}